\documentclass[a4paper, amsfonts, amssymb, amsmath]{article}
\usepackage[utf8]{inputenc}
\usepackage{color}
\usepackage[dvipsnames]{xcolor}
\usepackage{xfrac}
\usepackage{appendix}
\usepackage{soul}
\usepackage{ulem}
\usepackage{graphicx}
\usepackage{authblk}
\usepackage{amsmath,amssymb,amsfonts,graphicx}
\usepackage[a4paper,top=3cm,bottom=3cm,left=2.5cm,right=2.5cm]{geometry}
\usepackage{tikz}
\usepackage[pdftex, pdftitle={Article}, pdfauthor={Author}]{hyperref}
\usepackage{lscape}
\usepackage{relsize}
\usepackage{cite}

\begin{document}

\begin{titlepage}

\vspace{5cm}

\begin{center}

{\bf\Large Do heavy Monopoles hide from us ?}

\vspace{1cm} Huner Fanchiotti and C.A.~Garc\'\i a Canal \vskip 0.7cm

{\it IFLP(CONICET) and Department of Physics, University of La Plata}

{ \it C.C. 67 1900, La Plata, Argentina}\vskip 0.5cm

and \vskip 0.5cm

Vicente Vento \vskip 0.7cm

{\it  Departamento de F\'{\i}sica Te\'orica and Instituto de
F\'{\i}sica Corpuscular}

{\it Universidad de Valencia - Consejo Superior de Investigaciones
Cient\'{\i}ficas}

{\it 46100 Burjassot (Val\`encia), Spain, }

{\small Email: vicente.vento@uv.es}

\date{\today}

\end{center}
\vskip 1cm \centerline{\bf Abstract}

Dirac demonstrated that the existence of a single magnetic monopole in the universe could explain the discrete nature of electric charge. 
Magnetic monopoles naturally arise in most grand unified theories. However, the extensive experimental searches conducted thus far 
have not been successful. Here, we propose a mechanism in which magnetic monopoles bind deeply with neutral states, effectively 
hiding some of the properties of free monopoles. We explore various scenarios for these systems and analyze their detectability. 
In particular, one scenario is especially interesting, as it predicts a light state$-$an analog of an electron but with magnetic charge 
instead of electric charge$-$which we refer to as a \textit{magnetron}.

\vspace{2cm}

\noindent Pacs: 14.80.Hv, 95.30.Cq, 98.70.-f, 98.80.-k


\end{titlepage}

\section{Introduction}

The theoretical justification for the existence of classical
magnetic poles, hereafter called monopoles, is that they add
symmetry to Maxwell's equations and explain charge quantization
\cite{dirac1}. Dirac showed that the mere existence of a monopole
in the universe could offer an explanation of the discrete nature
of the electric charge. His analysis leads to the so-called Dirac
Quantization Condition (DQC), which in natural units reads,

\begin{equation}
e g = \frac{N}{2}, \quad \text{$N$ = 1, 2, ...} ;,
\label{DQC}
\end{equation}

\noindent where $e$ is the electron charge and $g$ the monopole
charge \cite{dirac1,Dirac:1948um}.

Monopoles have been described in different manners, either as point-like particles
or as topologically stable solutions of non-Abelian gauge theories \cite{book,kibble,'tHooft:1974qc,Polyakov:1974ek,preskill}.
For the purposes of our present investigation, their structure is irrelevant, and we shall consider them
as point-like particles endowed with a mass $M_m$ and a magnetic charge $g$ satisfying Eq.~(\ref{DQC}).

At present, despite intense experimental searches, there is no evidence of their existence
\cite{book,giacomelli,review,experiment,Abulencia:2005hb,Aad:2012qi,Aad:2015kta,MoEDAL:2016jlb,Acharya:2016ukt,MoEDAL:2021vix}. This state of
affairs has led us to investigate a possible mechanism by which monopoles could exist, hiding in a monopole-antimonopole bound state, monopolium~\cite{Vento:2007vy,Epele:2007ic}.
Here, we analyze a different mechanism. We assume that monopoles, at some point in the past, were bound to neutral particles, which we endow with a magnetic moment, and we study the properties of the bound system. The resulting states, named \textit{hideons}, are characterized by a magnetic charge and a magnetic moment. In the following sections, we study the dynamics of such a system for different scenarios. A particularly interesting one leads to a light state$-$an analog of an electron but with magnetic instead of electric charge$-$which we call \textit{magnetron}.

\section{Dynamics of the Hideon}

We study the coupling of a monopole, which we consider to be spin-0, and a neutral spin-1/2 particle with a magnetic moment. For the time being, we assume non-relativistic motion, which simplifies the treatment and is sufficient to demonstrate a mechanism by which two heavy particles can produce a light state. In this approximation, the bound state equation becomes~\cite{Sivers:1970zm,Olaussen:1983qc,Bracci:1983fe}

 \begin{equation}
- \frac{1}{2m}\left( \nabla^2 - \vec{\mu}\cdot \vec{B}\right) \Psi =E \Psi.
\label{Sch}
\end{equation}
Here $m=\frac{M_n M_m}{M_n+M_m}$ , $M_n$ and $M_m$ being the corresponding masses of the neutral and the monopole, $\vec{\mu}= g_a\frac{e}{2 M_n}\vec{S}$, where $g_a$ is the anomalous gyromagnetic ratio and $\vec{S}$ the spin of the neutral. The magnetic field created by the monopole is $\vec{B} = \frac{g \vec{r}}{r^3}$, where $g$ is the monopole magnetic charge. Substituting these definitions in Eq.(\ref{Sch}) we obtain,

 \begin{equation}
- \frac{1}{2m}\left( \nabla^2 - \frac{g_a}{4 M_n}\frac{\vec{S}\cdot \hat{r}}{r^2}\right) \Psi =E \Psi,
\end{equation}
where we have used the Dirac quantization condition $e g =\frac{1}{2}$. Note that the coupling is proportional to $e g$, and  thus not very large.

The operator $\vec{S}\cdot\hat{r}$ requires a very difficult diagonalization as can be seen easily from the equation

\begin{equation}
\vec{S}\cdot\hat{r}  = 2 \sqrt{\frac{2 \pi}{3}}\left( S_+ Y^1_{-1} (\theta,\varphi)+ S_z Y^1_0  (\theta,\varphi)- S_-Y^1_{+1} (\theta,\varphi)\right),
\end{equation}
where $S_\pm =S_x\pm i S_y$ and $Y ^l_m (\theta,\varphi)$ are the spherical harmonics.  Recalling now the property of the spherical harmonics

\begin{equation}
Y^1_{m_1} (\theta,\varphi) Y^l_{m_2} (\theta,\varphi) = \sum_{L=|l-1|}^{l+1} \sum_{m=-L}^L (1  l  0 0/ L 0) (1 l  m_1 m_2/ L m) Y^L_m (\theta,\varphi),
\end{equation}
where $(l_1 l_2 m_1 m_2/ l m)$ represent Clebsch-Gordan coefficients, we see that the angular-spin structure of the exact wave function is very complicated. We did not find an exact solution. To simplify the calculation we  will try as a variational ansatz  a  polarized wave function,

\begin{equation}
\Psi(r,\theta,\varphi) \chi_{\frac{1}{2} \; + \frac{1}{2}},
\label{ansatz}
\end{equation}
where $\chi_{\frac{1}{2} \; + \frac{1}{2}}$ is the corresponding spinor with projection about the axis $+ \frac{1}{2}$.  

Then

\begin{equation}
\begin{split}
&< \Psi(r,\theta,\varphi) \chi_{\frac{1}{2} \; + \frac{1}{2}}
| -S_-Y^1_{+1} + S_z Y^1_0 + S_+ Y^1_{-1} |\Psi(r,\theta,\varphi) \chi_{\frac{1}{2} \; + \frac{1}{2}}> \\
& = <\Psi(r,\theta,\varphi) \chi_{\frac{1}{2} \; + \frac{1}{2}}
|S_z Y^1_0 |\Psi(r,\theta,\varphi) \chi_{\frac{1}{2} \; + \frac{1}{2}}>. 
\end{split}
\end{equation}
With the ansatz Eq.(6) we are assuming that the neutral particle is polarized in a
certain direction. We could introduce a more general ansatz of a non-polarized neutral
using the density matrix formalism. This ansatz would require additional parameters
and complicate enormously the calculation without changing the aim of our presentation.

Thus from the point of view of the variational principle with a wave function of the above structure, Eq. (\ref{ansatz}), the hamiltonian of our problem, for the ground state, in spherical coordinates after eliminating the $\varphi$ dependence, turns out to be
 
 \begin{equation}
H= -\frac{1}{2m} \left(\frac{\partial^2}{\partial r^2} + \frac{2}{r}\frac{\partial}{\partial r}  + \frac{1}{r^2} \frac{\cos{\theta}}{\sin{\theta}} \frac{\partial}{\partial \theta} + \frac{1}{r^2}  \frac{\partial^2}{\partial \theta^2}\right) - \frac{g_a}{4 M_n} \frac{\cos{\theta}}{r^2}.
\label{hamiltonian}
\end{equation}
We have taken for the discussion $g_a$ positive. For a negative gyromagnetic ratio we should take the spinor polarization in the opposite direction.  From now on we shall use  $\gamma = \frac{g_a}{8M_n}$ always positive recalling that the spin polarization is associated with the sign of the gyromagnetic ratio. The spherical coordinates are defined to have the polarization axis at  $\theta=0$.

The effective magnetic moment-monopole potential leads to a singular hamiltonian since it falls at short distances like $1/r^2$.  In order to avoid the singularity we add a cutoff to the potential in line with the work of Schiff and Goebels \cite{Schiff:1967awg,Goebel} leading to,

\begin{equation}
V(r, \theta) = -\gamma  \frac{\cos{\theta} (1-e^{- \frac{r}{r_c}})}{r^2}
\end{equation}
where $r_c$, the cutoff radius, is a parameter whose possible values we shall discuss later. With this exponential factor the behavior of the potential as $r \rightarrow 0$ becomes $V(r,z)\rightarrow -  \gamma \frac{ z}{r r_c}$, which is Coulomb type and therefore leads to a non-singular hamiltonian, which becomes

 \begin{equation}
H= -\frac{1}{2m} \left(\frac{\partial^2}{\partial r^2} + \frac{2}{r}\frac{\partial}{\partial r}  + \frac{1}{r^2} \frac{\cos{\theta}}{\sin{\theta}} \frac{\partial}{\partial \theta} + \frac{1}{r^2}  \frac{\partial^2}{\partial \theta^2}\right) - \gamma \frac{\cos{\theta}\left(1-e^{-\frac{r}{r_c}}\right)}{r^2}.
\label{hamiltonian}
\end{equation}

Following the space-spin structure of  Eq. (\ref{ansatz}), we propose the following variational ansatz for the spacial part,

\begin{eqnarray}
r< r_c & \Psi_{<} (r, \theta) = (\alpha + \beta \cos{\theta}) \mathlarger{\frac{e^{-k r}}{r_c}} , \label{<}\\
r> r_c & \Psi_{>} (r, \theta) = (\alpha + \beta \cos{\theta}) \mathlarger{\frac{e^{-k r}}{r}} \label{>} ,
\end{eqnarray}
where $\alpha$ and $\beta$ are parameters to be determined and $k= \sqrt{2 m |E|}$.

The Coulombic behavior of the potential at short distances inspires the radial behavior of Eq.~(\ref{<}), while the long-range behavior of the interaction inspires the radial behavior of Eq.~(\ref{>}) \cite{bean}. The ansatz wave function is continuous, but it does not have a continuous derivative. To obtain the mass of the hideon, we will calculate the expectation value of the Hamiltonian and determine the values of $\alpha$ and $\beta$ by minimization. It is important to realize that if the Hamiltonian is bounded from below, the variational principle states that any variational ansatz will lead to an upper bound to the energy \cite{Landau:1991wop}. Thus, the corresponding result will provide an upper value for the binding energy, which is sufficient for our aim of proving that light states can be obtained from heavy particles.

The normalization of  the wave function establishes a relation between $\alpha$ and $\beta$ given by

\begin{equation}
\frac{|\beta|^2 + 3 |\alpha|^2 }{3}= \frac{2 k^3 r_c^2}{1- e^{-2 k r_c}(1+2 k r_c)}.
\label{normalization}
\end{equation}
The expectation value of the hamiltonian becomes

\begin{eqnarray}
2 m <H>& = & \frac{|\beta|^2 + 3|\alpha|^2}{3} \frac{1- e^{-2 k r_c}(1+ 2 k r_c- 2 k^2 r_c^2)}{2 k r_c^2} + \frac{2 |\beta|^2}{3} \frac{1-e^{-2 k r_c}}{k r_c^2} \nonumber \\
&& - \frac{|\beta|^2 + 3|\alpha|^2}{3} k e^{- 2 k r_c} + \frac{4 |\beta|^2}{3}\left( \frac{e^{- 2 k r_c}}{r_c} - 2 k \Gamma(0, 2 k r_c)\right)\nonumber \\
&&-\frac{2 m (\alpha^* \beta + \alpha \beta^*) \gamma}{3 r_c^2} \left(\frac{e^{-2 k r_c}}{k} - \frac{e^{-2 k r_c-1}}{k+ \frac{1}{2r_c} } \right)\nonumber \\
&&  - \frac{4 m (\alpha^* \beta + \alpha \beta^*) \gamma}{3} \left( \frac{e^{- 2 k r_c}}{r_c} - 2 k \Gamma(0, 2 k r_c) - \frac{e^{- 2 k r_c-1}}{r_c} + (2 k + \frac{1}{r_c}) \Gamma(0, 2 k r_c+ 1)\right)\nonumber\\
&&
\label{<H>}
\end{eqnarray}
Where $\Gamma$ is the incomplete gamma function.

We use from now on the reduced mass of the system $m$ as our energy scale, thus $\alpha^2 = \tilde{\alpha}^2 m$, $ \beta^2 = \tilde{\beta}^2 m$, $\gamma = \mathlarger{\frac{\tilde{\gamma}}{m} }$ , $k = \tilde{k} m$, $r_c = \mathlarger{\frac{\tilde{r}_c}{m}}$ , $E = \tilde{E} m = -\mathlarger{\frac{k^2}{2m}} = - \mathlarger{\frac{\tilde{k}^2}{2}} m$. With these substitutions the dimensions in the equations disappear since all energies are measured in units of $m$  and distances in units of $\frac{1}{m}$. Assuming $\tilde{\alpha}$ and $\tilde{\beta}$ real and omitting the tildes from now on the equation that determines  the energy becomes

\begin{eqnarray}
- k^2 & = & \frac{\beta^2 + 3\alpha^2}{3} \frac{1- e^{-2 k r_c}(1+ 2 k r_c- 2 k^2 r_c^2)}{2 k r_c^2} + \frac{2 \beta^2}{3} \frac{1-e^{-2 k r_c}}{k r_c^2} \nonumber \\
&& - \frac{\beta^2 + 3\alpha^2}{3} k e^{- 2 k r_c} + \frac{4 \beta^2}{3}\left( \frac{e^{- 2 k r_c}}{r_c} - 2 k \Gamma(0, 2 k r_c)\right)\nonumber \\
&&-\frac{4 \alpha \beta \gamma}{3 r_c^2} \left(\frac{e^{-2 k r_c}}{k} -\frac{e^{-2 k r_c-1}}{k+ \frac{1}{2r_c} } \right)\nonumber \\
&&  - \frac{8  \alpha \beta \gamma}{3} \left( \frac{e^{- 2 k r_c}}{r_c} - 2 k \Gamma(0, 2 k r_c) - \frac{e^{- 2 k r_c+1}}{r_c} + (2 k + \frac{1}{r_c}) \Gamma(0, 2 k r_c+ 1)\right)
\label{Energyeq}
\end{eqnarray}
subject to the condition
\begin{equation}
\frac{\beta^2 + 3 \alpha^2 }{3}= \frac{2 k^3 r_c^2}{1- e^{-2 k r_c}(1+2 k r_c)}.
\label{dimensionlessnormalization}
\end{equation}

These dimensionless equations are functions of $k r_c$, therefore the energy as a function of $r_c$ will correspond to $ k r_c = constant$. Thus our next job is to calculate the corresponding constant for different values of the potential characterized by $\gamma$.

\begin{figure}[htb]
\begin{center}
\includegraphics[scale= 0.9]{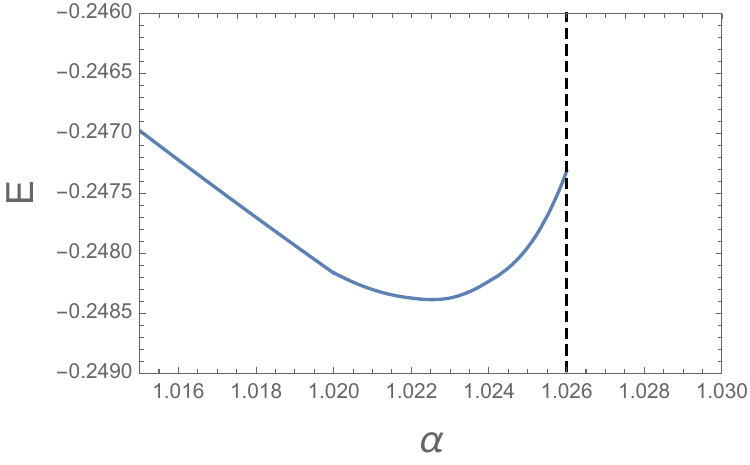}
\end{center}
\caption{We show the value of the binding energy of the system as a function of $\alpha$. Beyond 1.026 the energy equation Eq.(\ref{Energyeq}) has no solution. }
\label{minimum}
\end{figure}

 \section{Hideon structure}

The physical parameters in our calculations are $\gamma$ and $r_c$.  Let us discuss their possible values. We foresee two scenarios:

i) Light neutral scenario: $M_m >> M_n \rightarrow m=M_n$;

ii) Heavy neutral scenario: $M_n \sim M_m=M\rightarrow m=\frac{M}{2}$.

\noindent The third possibility, $M_n>> M_m$, is not considered because the magnetic moment is extremely small. 

In the first scenario $\gamma = \frac{g_a}{8}$, while in the second  $\gamma =\frac{g_a}{16}$. The value of $r_c$ becomes quite important, as the smaller $r_c$ is, the larger the contribution of the magnetic moment interaction. We treat $r_c$ as a parameter. Thus the values  we are going to investigate are $\gamma < 1$ which implies for the heavy neutral case $g_a < 16$ and for the light case $g_a < 8$, and in both cases we limit the value of $r_c <1$, which implies sizes smaller then $\frac{1}{m}$. It must be recalled that $m$ is the reduced mass which in the light neutral case is the light neutral mass, much smaller than the monopole mass, and in the heavy neutral case it is large, of the order of the monopole mass, and therefore the size of the system in this scenario will be very small.

Once we have fixed the magnitude  of these two dimensionless parameters we proceed to calculate the energy using Eqs.(\ref{Energyeq}) and (\ref{dimensionlessnormalization}) varying $\alpha$ looking for minima. In Fig. \ref{minimum} we show the minimization procedure for $\gamma =1$ and $r_c=0.5$. As seen in the figure, for certain values of $\alpha$, $\alpha> 1.026$, Eq. (\ref{Energyeq}) ceases to be satisfied. We take the minimum value as our best value recalling  that it is an upper limit of the true binding energy. 

\begin{figure}[htb]
\begin{center}
\includegraphics[scale= 0.8]{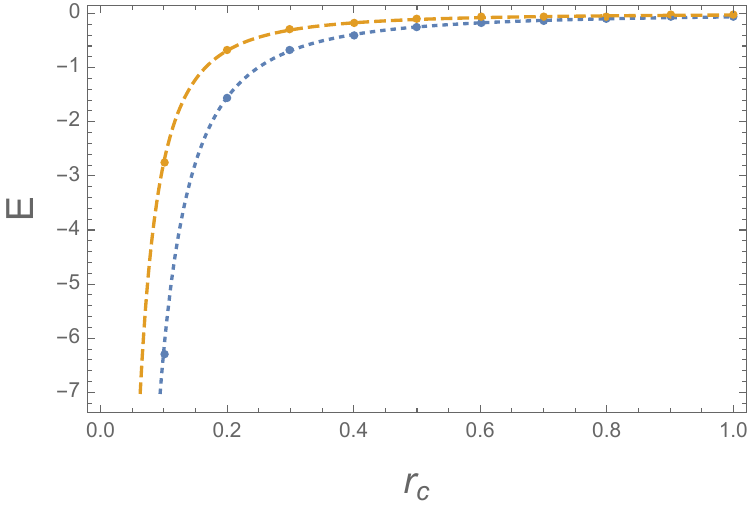} 
 \vskip -6.cm
 \hskip  3.5cm
\includegraphics[scale= 0.5]{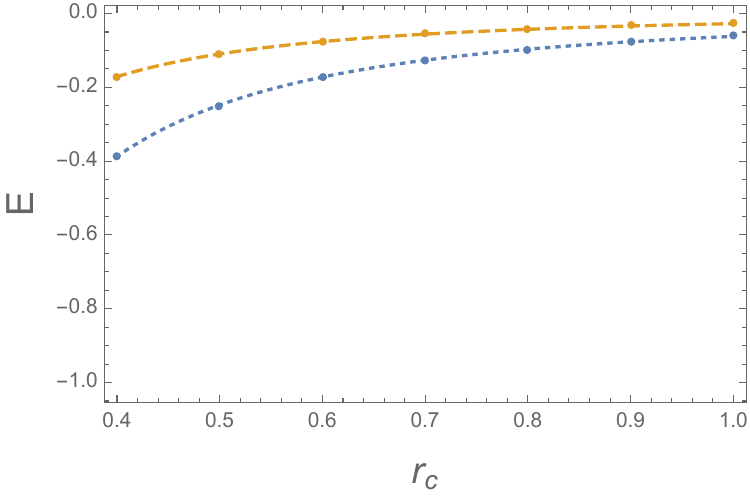} 
\end{center}
\vskip 2cm
\caption{We show the value of the binding energy of the system as a function of $r_c$. The  inset shows the behavior for large $r_c$. The curves represent $k r_c$ = constant. The dotted line corresponds to $\gamma=1$ and $k r_c =0.352$ and the dashed line to $\gamma= 0.5$ and $k r_c=0.234$.}
\label{Energyfig}
\end{figure}

In Fig. \ref{Energyfig} we show the binding energy as a function of $r_c$ for two values of $\gamma$, $0.5$ and $1$. We note that the binding energy depends crucially on the cut-off radius $r_c$. It is important to emphasize that the Hamiltonian without a cut-off is unbounded from below. In the case of the $\frac{1}{r^2}$ potential the elimination of the  the cut-off has lead, in order to obtain physical results, to a sort of {\it renormalization} process that requires experimental input \cite{Case:1950an,bean,Bouaziz:2017ijz}. Since in our case we lack experimental information we treat the cut-off $r_c$ as physical by assuming that the monopoles have a finite size.  For our variational wave function the relation between the energy and the cut-off radius is very simple $k r_c = constant$ as discussed earlier. In the figure we show a few numerical points which we have used to calculate the constant, along with the final curves  for $\gamma=1$ with $ k r_c= 0.352$ and for $\gamma = 0.5$ with $k r_c= 0.234$. The {\it renormalization} process, discussed in refs.\cite{Case:1950an,bean,Bouaziz:2017ijz}, which would eliminate the $r_c$ dependence, requires the knowledge of a hideon observable.

\begin{figure}[htb]
\begin{center}
\includegraphics[scale= 0.8]{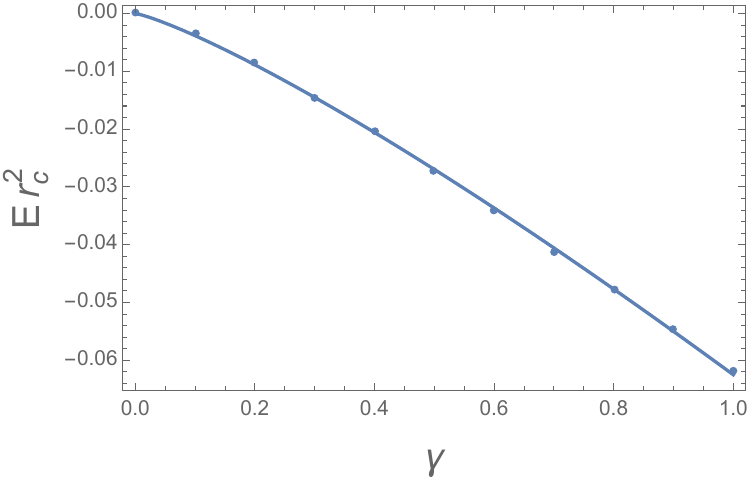}
\end{center}
\caption{We show the relation between the potential strength $\gamma$, the binding energy $E$ and the cut-off radius $r_c$. Note that for each gamma we obtain an equation $E r_c^2 = \mbox{constant}$ .}
\label{Energyfig2}
\end{figure}

In Fig. \ref{Energyfig2} we show the value of $E r_c^2$ for different values of $\gamma$. The corresponding relation  can be approximated by 

\begin{equation}
E r_c^2 \sim -0.06251 \gamma^{1.21239}.
\end{equation}
We note that for any value of $\gamma$ the binding energy can become large for small $r_c$. This implies that monopoles with small $r_c$ produce very large binding energies and therefore might lead to  light hideons as we shall discuss.

\begin{figure}[htb]
\begin{center}
\includegraphics[scale= 0.8]{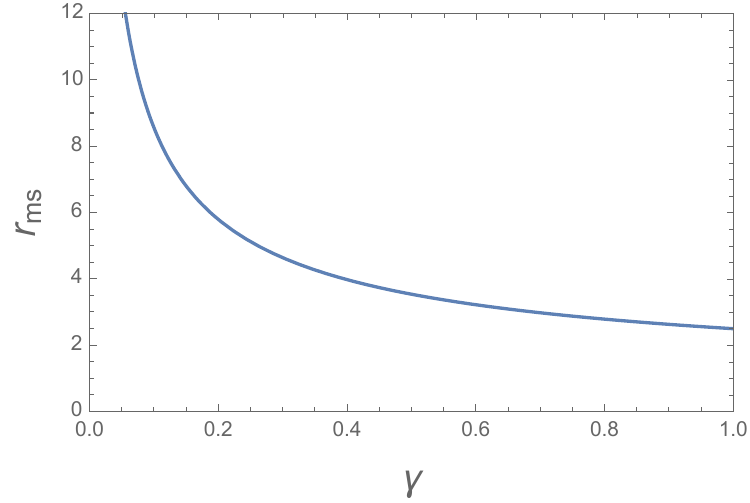} 
\end{center}
\caption{We show the relation between the potential strength $\gamma$ and the mean square radius $r_{ms}$. Note that  Eq.(\ref{rms}) does depend on $\gamma$ only through $kr_c$ . Thus different states lead to the same $r_{ms}$ since $k r_c$ is constant for a value of $\gamma$.}
\label{meansquareradius}
\end{figure}

Given the fact that $k r_c = \mbox{constant}$ for every potential strength, the observables determined from the wave function are functions only of $k r_c$ and therefore their values will be constant for these observables for different values of $r_c$. For example the mean square radius comes out to be in units of $1/m$

\begin{equation}
r_{ms}^2 = <\psi |r^2| \psi> = \frac{3 - e^{-2 k r_c}(3+ 6 k r_c + 5 k^2 r_c^2 + 2 k^3 r_c^3)}{(1- e^{-2 k r_c}(1+2 kr_c))k^2 r_c^2}.
\label{rms}
\end{equation}
We plot the corresponding values in Fig. \ref{meansquareradius} for different values of $\gamma$.

 After performing this non-relativistic calculation, one can analyze the details of the dynamics and realize that the strongest bound states should be described by a relativistic equation. However, the emphasis of this paper is not on the precise calculation of the possible masses of these states, but on demonstrating that, due to the radial structure of the magnetic moment interaction, very deeply bound states might arise. To describe these states precisely, one would need to solve the Dirac equation~\cite{Semenov}, which would greatly increase the complexity of the calculation without providing any new physical insights at this stage.

In any case, to confirm the above statements, we have performed an approximate relativistic calculation where we have used the relativistic kinetic term~\cite{Dixit}.

\begin{figure}[htb]
\begin{center}
\includegraphics[scale= 0.6]{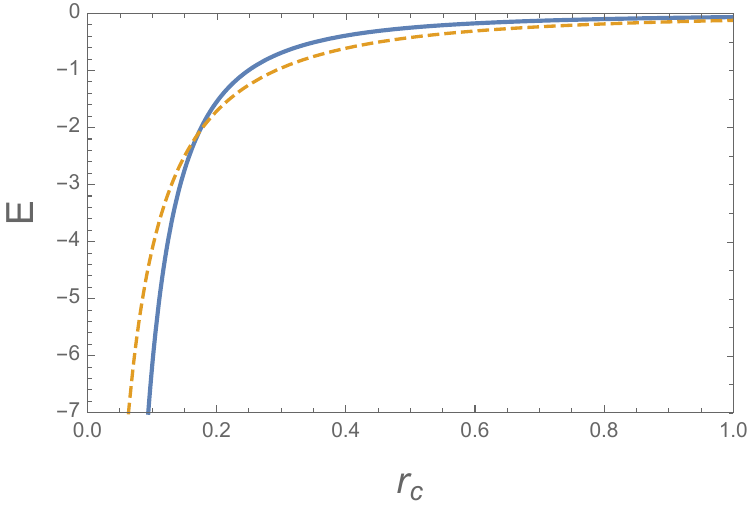} \hspace{0.5cm}\includegraphics[scale= 0.6]{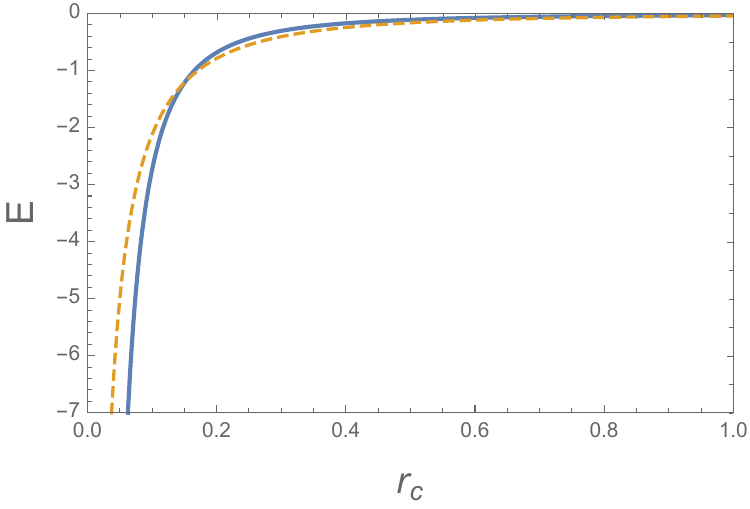} 
\end{center}
\caption{We show the relation between the binding energy $E$ and the core radius $r_c$ for the non-relativistic (solid) and the relativistic (dashed) approximations  for $\gamma= 1$ (left panel) and $\gamma=0.5$ (right panel). In both cases the two approximations differ more as we increase the strength of the potential.}
\label{Energyfigrel}
\end{figure}

\begin{equation}
\frac{p^2}{2m} \rightarrow  \sqrt{p^2+m^2}-m.
\end{equation}

\noindent Using  the same variational ansatz we have substituted 

\begin{equation}
<\Psi| \sqrt{p^2+m^2}-m|\Psi> \rightarrow \sqrt{<\Psi|p^2+m^2|\Psi> - m}.
\end{equation}

\noindent This is equivalent to use the relativistic expansion and perform the approximation in the expectation values as
\begin{equation}
<\Psi| \sqrt{p^2+m^2}-m|\Psi>\sim <\Psi|\frac{p^2}{2m} -  \frac{p^4}{8m^3} +  \frac{p^6}{16m^5}  - \ldots|\Psi> \rightarrow \frac{ <\Psi|p^2|\Psi>}{2m} -  \frac{<\Psi|p^2|\Psi>^2}{8m^3} +  \frac{<\Psi|p^2|\Psi>^3}{16m^5}  -\ldots.
\end{equation}

\noindent This approximation corresponds to use $|\Psi>$ as a complete set, i.e. the so called closure approximation,

\begin{align}
<\Psi|p^4|\Psi> &\rightarrow <\Psi|p^2|\Psi> <\Psi|p^2|\Psi>  \nonumber \\
<\Psi|p^6|\Psi> &\rightarrow <\Psi|p^2|\Psi> <\Psi|p^2|\Psi>  <\Psi|p^2|\Psi> \nonumber \\
\ldots
\end{align}

It is clear that this is not a very powerful approximation but it will give us a hint of how the above results shown in Fig.\ref{Energyfig} might be modified. The comparison is shown in Fig. \ref{Energyfigrel}. We note that physically the two calculations behave qualitatively very similar, i.e. both  allow for very strong bound states, and in both the dynamics governing the depth of the bound state is associated fundamentally with the core radius $r_c$. The relativistic calculation is less binding for small $r_c$. For large $r_c$ both calculations are almost identical.

In what follows we will perform a qualitative analysis using the results of the non relativistic calculation analyzing the possible scenarios that appear taking into account the different possible values of the masses of the monopoles and the neutral.

\section{Hideon properties and detectability}

In order to study the properties of the hideon states let us recall the two scenarios mentioned above:

i)  Light neutral scenario: $M_m >> M_n \rightarrow M_n \sim m$.

ii) Heavy neutral scenario: $M_m \sim M_n = M \rightarrow m \sim \frac{M}{2} $.

i)In the first case, the binding energy is small for conventional values of $r_c$ and $\gamma$, since the outcome is in units of the reduced mass $m$, which is essentially the mass of the light fermion. The hideon then behaves as a heavy fermion with magnetic charge, magnetic moment, and a large mass $\sim M_m$. As an example, one such very light fermion could be the neutron. In this case, the hideon will appear as a heavy particle with a large magnetic charge and a measurable magnetic moment. If the neutral particle is heavier than the neutron, then the magnetic moment will be small, $\mu \sim \frac{1}{M_n}$.

ii) In the second case, by an appropriate choice of parameters, the binding energy can be  made large, namely for large  $\gamma$  and/or small $r_c$. For example, one can achieve, in our approximation, a zero mass hideon, i.e. a binding energy of $-4m$, by choosing   $g_a \sim 4-6$, that is  $\gamma \sim 0.25 -0.4$, and $r_c \sim 0.05-0.07$, as can be obtained from Fig. \ref{Energyfig2}. Note that the $\gamma$ associated with a neutron is $0.24$. Therefore very light hideons might exist. These hideons are light particles with a huge magnetic charge and small magnetic moment, which behave as a magnetic electron, therefore we call them  magnetrons. In the case the binding energy is small, i.e. smaller values of $\gamma$ and larger $r_c$,  these hideons behave as in i) with the difference that their magnetic moment will be very small since $\mu \sim \frac{1}{M_n}\sim \frac{1}{M_m}$.

Let us show some numerical examples. Assume a GUT monopole, $M_m\sim10^{16}$ GeV,  with a GUT neutral leading to $\gamma=0.25$, which corresponds to a magnetic moment of $ \sim10^{-19}$ Bohr magnetons, $r_{ms} \sim10^{-16} fm$,  and to $Er_c^2 \sim-0.01$, which for large $r_c\sim1$ leads to a binding energy of $ \sim-10^{14} GeV$. Thus the mass of the hideon  is a GUT mass $\sim2\;10^{16}$ GeV. Given the large magnetic charge, a binding energy much smaller than its mass, and the small magnetic moment, the heavy hideon would appear experimentally as a GUT monopole, since the effect of the magnetic moment  would be small in the detectors.  

 Let us assume now a low mass monopole , $M_m\sim10^{3}$ GeV,  with a similar mass neutral leading to $\gamma=0.25$, which corresponds to a magnetic moment of $\sim10^{-5}$ Bohr magnetons, $r_{ms}\sim10^{-3} fm$, and to $Er_c^2\sim-0.01$, which for large $r_c\sim1$ leads to a binding energy of $\sim-10$ GeV. Thus the mass of the hideon  is  low $\sim2\;10^{3}$ GeV. Given the large magnetic charge, the small binding energy, and the small magnetic moment, this hideon will appear experimentally as a monopole, unless it breaks up in the detector, since the binding energy is relatively small for high energy collisions. Detection will be then notably different, since in the final state one has a monopole and a neutral. The neutral will be most probably undetected given the small magnetic moment, but it will carry momentum and energy. This will show up as a monopole with an irregular trajectory, signaling the previous existence of a hideon. In a Moedal type detector  \cite{MoEDAL:2016jlb}, one would get probably  two imprints in the detector, a highly ionising one from the monopole, and less ionizing one from the neutral. Trajectory reconstruction might determine the bound state nature of the initial state.
 
 In case i) if the neutral is very light, i.e. a neutron, the result would be similar to that described above except that the magnetic moment would be measurable.
 
Let us now turn to the light hideon scenario which is characterized by the existence of a new particle since, given the huge binding energy, the constituents are almost confined. This new particle, the magnetron, behaves like a magnetic electron, with a very large magnetic charge and a small magnetic moment. Thus it behaves as a light fermionic monopole. 

 Let us assume a constituent GUT monopole, $M_m\sim10^{16}$ GeV,  with a GUT neutral leading to $\gamma=0.5$, which corresponds to a magnetic moment of $\sim10^{-19}$ Bohr magnetons, $r_{ms}\sim10^{-16} fm$, and to $Er_c^2\sim-0.02$, which for small $r_c\sim0.11$ leads to a binding energy of $\sim-2 \,10^{16}$ GeV. Thus the mass of the magnetron  can be very small by the proper choice of $\gamma$ and $r_c$. This magnetron will manifest itself like a light monopole since the magnetic moment effect is small. Given its massive constituents it cannot be produced in accelerators.

 Let us assume a low mass monopole, $m\sim10^{3}$ GeV,  with a similar mass neutral leading to $\gamma=0.5$, which corresponds to a magnetic moment of $\sim10^{-5}$ Bohr magnetons, $r_{ms} \sim 10^{-3} fm$, and to $Er_c^2 \sim-0.02$, which for small $r_c\sim0.11$ leads to a binding energy of  $\sim-2\,{10^3}$ GeV. Thus the mass of the magnetron  can be very small by the proper choice of $\gamma$ and $r_c$. This hideon will manifest itself like a very light fermionic monopole since the magnetic moment effect is small. However, it can kinematically be produced in accelerators, and in the presence of very strong magnetic fields the magnetic moment might show up.

 \section{Conclusions}

 Monopoles have been the subject of extensive study and experimental searches since Dirac revealed their quantum mechanical properties~\cite{Dirac:1948um}. The lack of success in detecting them experimentally has led physicists to propose that they may exist in an almost magnetically neutral bound state known as monopolium~\cite{Vento:2007vy,Epele:2007ic}. Certain experimental analyses have translated the absence of detection into mass limits in the TeV range~\cite{Abulencia:2005hb,Aad:2015kta,MoEDAL:2021vix}. Grand Unified Theories (GUT), on the other hand, place their mass limits at the GUT scale~\cite{'tHooft:1974qc,Polyakov:1974ek,preskill}. In this work, we study a mechanism that effectively hides monopoles within bound states. These states retain the monopole charge but alter their kinematical and electromagnetic properties by introducing a magnetic moment. Essentially, this mechanism transforms a bosonic monopole into a fermionic monopole, and in one particular scenario, the resulting state can be very light. Detecting such a bound state requires consideration of its mass (which could be very small), its large magnetic charge, and its small magnetic moment.
 
 Two scenarios emerge from our study. In the first scenario, the particles in the bound state are loosely bound, resulting in very massive \textit{hideons}. These states possess a large magnetic charge and a small magnetic moment, making them similar to monopoles from a detection perspective. However, if the bound state is broken upon detection, the kinematics would involve a two-particle final state. Whether the small magnetic moment affects their astrophysical behavior remains an open question.
 
 The second scenario is characterized by a large binding energy, leading to light states with both magnetic charge and magnetic moments. These fermionic states, which we refer to as \textit{magnetrons}, have a very large magnetic charge and a small magnetic moment. Due to their small mass, these states would be easier to detect. Moreover, if their constituents are in the TeV range, they could potentially be produced at colliders.

\section*{Acknowledgments}

HF and CAGC were partially supported by ANPCyT, Argentina. VV was supported by a PROMETEO/2021/083 grant from GVA and by a Severo Ochoa grant  CEX2023-001292-S from  MCIU/AEI.\\

\noindent {\bf Data Availability Statement:} No Data associated in the manuscript.


\begin{thebibliography}{99}

\bibitem{dirac1} P.A.M. Dirac, Proc. Roy. Soc. {\bf A133} (1931)
60, Phys. Rev. {\bf 74} (1940) 817.

\bibitem{Dirac:1948um}
  P.~A.~M.~Dirac,
  Phys.\ Rev.\  {\bf 74} (1948) 817.
  doi:10.1103/PhysRev.74.817

  
\bibitem{book} N. Craigie, G. Giacomelli, W. Nahern and Q.
Shafi,{\sl Theory and detection of magnetic monopoles in gauge
theories}, World Scientific, Singapore{1986}.


\bibitem{kibble} T.W.B. Kibble, J. Phys. {\bf A 9} (1976) 1387,
Phys. Rep. {\bf 67} (1980) 183; A. Vilenkin, Phys. Rep. {\bf 121}
(1985) 263.


\bibitem{'tHooft:1974qc}
  G.~'t Hooft,
  Nucl.\ Phys.\ B {\bf 79} (1974) 276.
  

  \bibitem{Polyakov:1974ek}
  A.~M.~Polyakov,
  JETP Lett.\  {\bf 20} (1974) 194
   [Pisma Zh.\ Eksp.\ Teor.\ Fiz.\  {\bf 20} (1974) 430].


\bibitem{preskill} J.P. Preskill, Phys. Rev. Lett. {\bf 43} (1979)
1365.

\bibitem{giacomelli} G. Giacomelli and L. Patrizii, hep-ex/0506014.


\bibitem{review} K. A. Milton, hep-ex/0602040.

\bibitem{experiment} Review of Particle Physics, S. Eidelman et
al. Phys. Lett. {\bf B592} {2004} 1.

\bibitem{Abulencia:2005hb}
  A.~Abulencia {\it et al.}  [CDF Collaboration],
  Phys.\ Rev.\ Lett.\  {\bf 96} (2006) 201801
  [hep-ex/0509015].


\bibitem{Aad:2012qi}
  G.~Aad {\it et al.}  [ATLAS Collaboration],
  Phys.\ Rev.\ Lett.\  {\bf 109} (2012) 261803
  [arXiv:1207.6411 [hep-ex]].

\bibitem{Aad:2015kta}
  G.~Aad {\it et al.} [ATLAS Collaboration],
  Phys.\ Rev.\ D {\bf 93} (2016) no.5,  052009
  doi:10.1103/PhysRevD.93.052009
  [arXiv:1509.08059 [hep-ex]].
  
\bibitem{MoEDAL:2016jlb}
  B.~Acharya {\it et al.} [MoEDAL Collaboration],
  JHEP {\bf 1608} (2016) 067
  doi:10.1007/JHEP08(2016)067
  [arXiv:1604.06645 [hep-ex]].
  

\bibitem{Acharya:2016ukt}
  B.~Acharya {\it et al.} [MoEDAL Collaboration],
  Phys.\ Rev.\ Lett.\  {\bf 118} (2017) no.6,  061801
  doi:10.1103/PhysRevLett.118.061801
  [arXiv:1611.06817 [hep-ex]].

\bibitem{MoEDAL:2021vix}
B.~Acharya \textit{et al.} [MoEDAL],
Nature \textbf{602} (2022) no.7895, 63-67
doi:10.1038/s41586-021-04298-1
[arXiv:2106.11933 [hep-ex]].

\bibitem{Vento:2007vy}
  V.~Vento,
  Int.\ J.\ Mod.\ Phys.\ A {\bf 23} (2008) 4023
  doi:10.1142/S0217751X08041669
  [arXiv:0709.0470 [astro-ph]].


\bibitem{Epele:2007ic}
  L.~N.~Epele, H.~Fanchiotti, C.~A.~Garcia Canal and V.~Vento,
  Eur.\ Phys.\ J.\ C {\bf 56} (2008) 87
  doi:10.1140/epjc/s10052-008-0628-0
  [hep-ph/0701133].

\bibitem{Sivers:1970zm}
D.~W.~Sivers,
Phys. Rev. D \textbf{2} (1970), 2048-2054
doi:10.1103/PhysRevD.2.2048

\bibitem{Olaussen:1983qc}
K.~Olaussen, H.~A.~Olsen, P.~Osland and I.~Overbo,
Nucl. Phys. B \textbf{228} (1983), 567-587
doi:10.1016/0550-3213(83)90560-6


\bibitem{Bracci:1983fe}
L.~Bracci and G.~Fiorentini,
Nucl. Phys. B \textbf{232} (1984), 236-262
doi:10.1016/0550-3213(84)90566-2

\bibitem{Schiff:1967awg}
L.~I.~Schiff,
Phys. Rev. \textbf{160} (1967) no.5, 1257-1262
doi:10.1103/physrev.160.1257

\bibitem{Goebel} 
C~J.~Goebel, Quanta, Essays in Theoretical Physics, eds. P.G.O. Freund, C.J. Goebel
and Y. Nambu (Chicago, 1970).


\bibitem{Landau:1991wop}
L.~D.~Landau and E.~M.~Lifshits,
Butterworth-Heinemann, 1991,
ISBN 978-0-7506-3539-4
doi:10.1016/C2013-0-02793-4



\bibitem{bean} S. R. Beane, P.F.  Bedaque, L. Childress, A. Kryjevski, A., J. McGuire, J. and U. van Kolck, Phys. Rev. A {\bf 64} 042103.


\bibitem{Case:1950an}
K.~M.~Case,
Phys. Rev. \textbf{80} (1950), 797-806
doi:10.1103/PhysRev.80.797


\bibitem{Bouaziz:2017ijz}
D.~Bouaziz and T.~Birkandan,
Annals Phys. \textbf{387} (2017), 62-74
doi:10.1016/j.aop.2017.10.004
[arXiv:1711.04158 [quant-ph]].

\bibitem{Semenov}
V.V.~Semenov 
J. Phys. A: Math. Gen. \textbf{23} (1990) L721

\bibitem{Dixit}
Anant Dixit, Yannick Hinschberger, Jens Zamanian, Giovanni Manfredi, and Paul-Antoine Hervieux
Physical Review A  \textbf{88} (2013) 032117 

\end{thebibliography}
\end{document}